\documentclass[traditabstract]{aa}
\usepackage{natbib}
\usepackage{graphicx}
\usepackage{txfonts}

\begin{document}

\title{The double degenerate LP~400-22 revisited}

\author{S. Vennes\inst{1,2}, A. Kawka\inst{1}, T.R. Vaccaro\inst{2,3}
, \and N.M. Silvestri\inst{4}
}
\institute{Astronomick\'y \'ustav AV \v{C}R, Fri\v{c}ova 298,
 CZ-251 65 Ond\v{r}ejov, Czech Republic; vennes, kawka@sunstel.asu.cas.cz.
 \and
 Visiting Astronomer, Kitt Peak National Observatory, National Optical Astronomy Observatory, which is operated by the Association of Universities for Research in Astronomy (AURA) under cooperative agreement with the National Science Foundation. 
 \and
Department of Physics \& Astronomy, Francis Marion University, Box 100547,
 Florence, SC 29501, USA; tvaccaro@fmarion.edu.
 \and
 Department of Astronomy, University of Washington, Box 351580, Seattle, WA 98195, USA; nms@astro.washington.edu.
} 
\date{Received ; accepted }

\abstract{
We re-examine the properties of the high-velocity and extremely low-mass 
white dwarf LP~400-22 and its close companion. Based on an extended observation timeline, 
we determined a binary period of $P=1.01016$ d, somewhat longer than the previously published 
period, and a mass function $f(M_2)= 0.180\,M_\odot$, implying a mass for the companion 
$M_2> 0.41\,M_\odot$. We also re-appraised the mass and cooling age of the white dwarf 
using low-metallicity ($Z=0.001$) evolutionary models appropriate for an old halo 
member, $M_1=0.19\,M_\odot$, and $t_{\rm cool}\approx1.8$ Gyr, and we infer a mass of $0.85-1.0 \,M_\odot$ for the progenitor of the white dwarf. We discuss the likely origin
of this system.
}


\titlerunning{Binary LP~400-22}
\authorrunning{Vennes et al.}

\maketitle

\section{Introduction}

LP~400-22 (NLTT~54331) is an extremely low-mass (ELM) white dwarf identified
by \citet{kaw2006}. This object is also remarkable for its peculiar kinematics characterised
by a high tangential velocity $\varv_{\rm tan}\approx 400$ km s$^{-1}$ \citep{kaw2006}
and for its membership in a $\sim 1$ day binary \citep{kil2009}. The companion still has to be identified but it
is most likely another white dwarf or a neutron star. Evolutionary scenarios
\citep[see][]{wil2002,nel2001} that produce low-mass white dwarfs
typically involve a common-envelope (CE) phase responsible for orbital shrinkage, while
the most massive component climbs toward the giant 
branch or, alternatively, while it reaches the asymptotic giant branch. 
The most massive component eventually collapses to form a neutron star or a white dwarf,
and, depending on the resulting orbital period and on the secondary mass, the system 
may experience episodes of Roche-lobe overflow (RLOF), while the secondary is still 
on the main-sequence
or, later, while it climbs the giant branch. Either path leads to the formation of a low-mass helium
white dwarf, with a total systemic age characteristic of the old disk or halo population for systems with a low
initial secondary mass, or significantly lower ages for systems with higher secondary masses.

Knowledge of the total age of the system is critical in helping estimate the likely 
mass and metallicity of the white dwarf progenitor, hence in justifying 
a particular choice of evolutionary models.
In this regard, a kinematical study of the remnant stars is useful in estimating the 
epoch of formation \citep{chi2000}. 

A few ELM white dwarfs ($M < 0.2\ M_\odot$) 
have been identified as companions to pulsars \citep{van1996,bas2006}.
In addition, the white dwarf SDSS~J0917$+$46, which was discovered as part of a 
search for hypervelocity B-type stars \citep{kil2007a}, is also in a $7.5936\pm0.0024$ h
binary \citep{kil2007b} with a companion still to be classified\footnote{Proceeding by elimination,
\cite{agu2009} argue in favour of a white dwarf companion.}. 

In an effort to improve our understanding of the binary LP~400-22, we obtained two series 
of intermediate resolution spectra (\S 2), which we use to refine the white dwarf 
atmospheric parameters (\S 3.1). Based on new and published radial velocity measurements, 
we constrain the binary ephemeris, and revise the kinematics of the system, as well as 
the mass-function for the unseen 
companion (\S 3.2). We also re-evaluate the parameters of the white dwarf 
based on low-metallicity evolutionary models appropriate for an old halo member (\S 3.3). Finally, we present our conclusions (\S 4).

\section{Spectroscopy}

Table~\ref{tbl-1} shows the log of observations. 
LP~400-22 was observed on 2006 September 30 using the dual imaging spectrograph (DIS) 
attached to the 3.5 m telescope at the Apache Point Observatory (APO).
We obtained six exposures of 30 minutes each. 
We used the 830.8 lines mm$^{-1}$ grating on the red side to 
obtain a spectral range of 6440 to 8150 \AA\ with a dispersion of 0.84 \AA\ pixel$^{-1}$, 
and the 1200 lines mm$^{-1}$ grating on the blue side to obtain a spectral range 
of 3830 to 5030 \AA\ with a  dispersion of 0.62  \AA\ pixel$^{-1}$.
The slit width was set at 1.5$\arcsec$ resulting in a resolution of 2.1 \AA\ in  
the red and 1.7 \AA\ in the blue.
We concurrently obtained a series of HeNeAr comparison arcs.
We corrected the wavelength scale for each science exposure using the HgI$\lambda 4358.335$ sky line on the blue side
and the OH$\lambda6533.05$ sky line on the red side.

\begin{figure}[t!]
\centering
\resizebox{\hsize}{!}{\includegraphics[width=0.95\columnwidth]{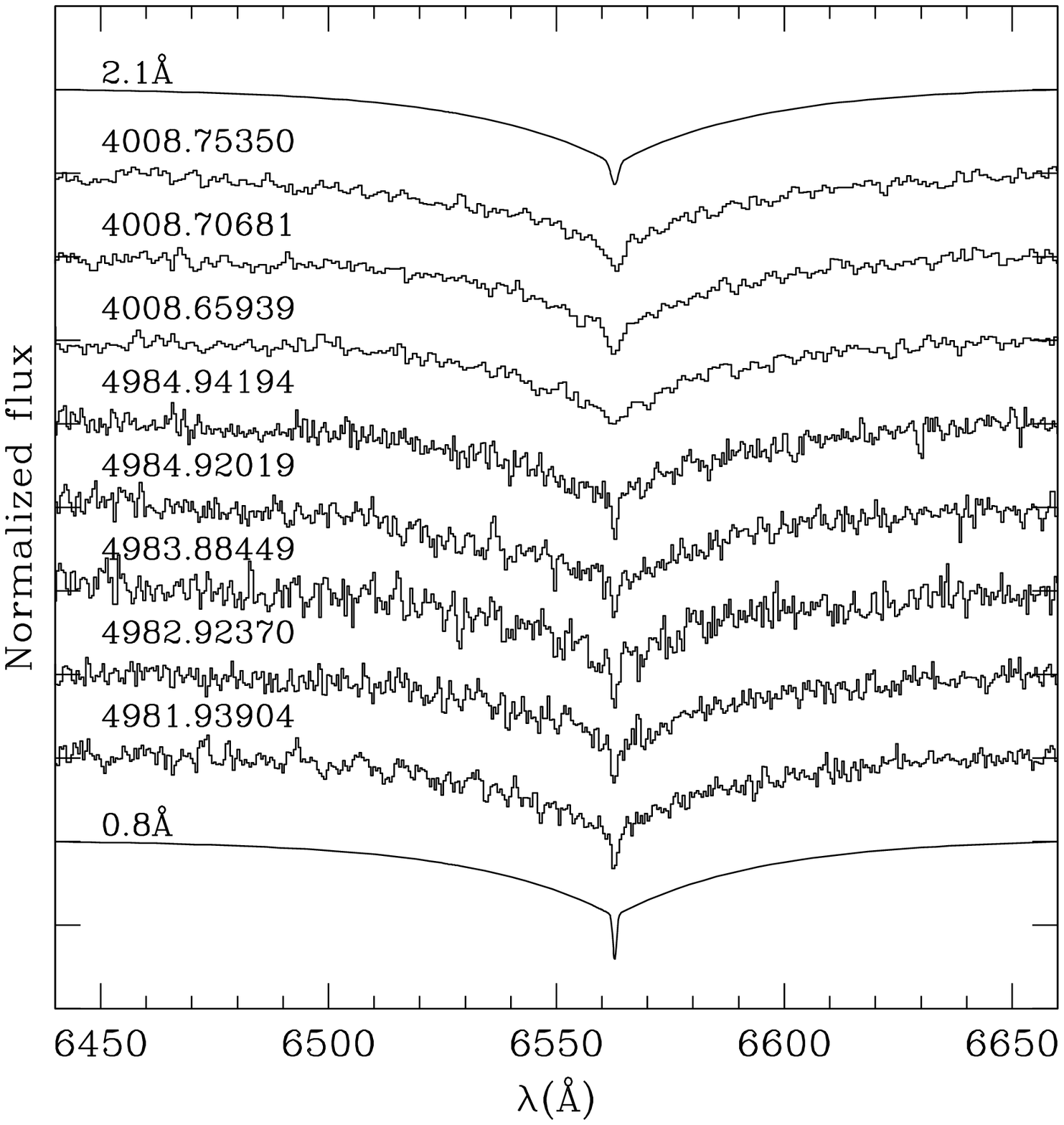}}
\caption{APO and KPNO spectra centred on H$\alpha$ (labelled with HJD) compared to synthetic spectra (labelled with spectral resolution) showing
a narrow photospheric absorption core.}
\label{fig1}
\end{figure}

\begin{table}
\caption{Observation log}
\label{tbl-1}
\centering
\begin{tabular}{ccccc}
\hline\hline
Observatory & Start UT date &  Time  & Object & $t_{\rm exp}(s)$   \\
\hline
APO  & 2006 Sep 30&  03:11:01 & LP~400-22& 1800    \\
     &             &  03:42:22  & HeNeAr  & 90 \\
     &            &  03:45:05 & LP~400-22& 1800    \\
     &             &  04:16:20  & HeNeAr  & 180\\
     &            &  04:20:44 & LP~400-22& 1800    \\
     &            &  04:51:57 & LP~400-22& 1800    \\
     &             &  05:23:09  & HeNeAr  & 180\\
     &            &  05:28:07 & LP~400-22& 1800    \\
     &            &  05:59:00 & LP~400-22& 1800    \\
     &             &  06:30:12  & HeNeAr  & 180\\
KPNO & 2009 May 30 &  10:18:41  & LP~400-22&  1800    \\
     &             &  10:50:38  & HeNeAr  & 30 \\
     & 2009 May 31 &  09:56:28  & LP~400-22&  1800    \\
     &             &  10:28:23  & HeNeAr  & 30 \\
     & 2009 May 01 &  08:59:53  & LP~400-22&  1800    \\
     &             &  09:32:08  & HeNeAr  & 30 \\
     & 2009 May 02 &  09:51:10  & LP~400-22&  1800    \\
     &             &  10:22:29  & LP~400-22&  1800    \\
     &             &  10:55:07  & HeNeAr  & 30 \\
\hline
\end{tabular}
\end{table}

Next, we observed LP~400-22 from 2009 May 29 to June 2 using the Ritchey-Chretien spectrograph attached to the 4 m telescope at Kitt Peak National Observatory (KPNO).
We used the KPC-24 grating (860 lines mm$^{-1}$) in the second order resulting in a
dispersion of 0.51 \AA\ pixel$^{-1}$ and a spectral range from 6230 to 6860 \AA. The slit width was set to 
1.5$\arcsec$, resulting in a resolution of $\approx0.8$\AA\ at H$\alpha$ and degrading to $\approx1.5$\AA\ on the fringes.
We also secured a concurrent series of HeNeAr comparison arcs.
To ascertain the stability of the wavelength scale, we also observed the radial velocity standard HD~112299 each night. We measured an average velocity and dispersion $\varv_{\rm helio}=4.6\pm1.5$ km~s$^{-1}$ close to 
published values \citep[$\varv_{\rm helio}=3.9\pm0.6$ km s$^{-1}$,][]{maz1996}.
Moreover, the radial velocity measurements of the sky line OH$\lambda6533.05$ 
demonstrate the stability of the wavelength scale with a standard deviation of only 0.7 km\,s$^{-1}$.

The spectra were flux-calibrated using the spectro-photometric standards Feige 110, Feige~34, and
HZ~44. Finally, all spectra were reduced using standard IRAF procedures.
Figure~\ref{fig1} shows the spectra near H$\alpha$. To improve the signal-to-noise ratio 
of the APO observations we co-added the spectra by pairs. The sharp H$\alpha$ absorption core proved to be a precise velocity marker in both science, particularly those obtained at KPNO, and velocity standard 
exposures.

Table~\ref{tbl-2} shows measured radial velocities and mid-exposure times. 
The velocities at KPNO were obtained by fitting Voigt profiles to the narrow H$\alpha$ absorption core using the IRAF `splot' routine. An average error was estimated by taking successive measurements varying the pseudo-continuum location. We measured the velocities at APO by calculating the average and
standard deviation of five independent measurements obtained by fitting Voigt profiles to (1) the H$\alpha$ line core and (2) the H$\gamma$ line core, and by (3) cross-correlating the spectra with a template between 6540 and 6580\AA, (4) between 4300 and 4400\AA, and (5) between 4325 and 4355 \AA. The red and blue templates were obtained by summing all available spectra of LP~400-22.

\section{Analysis}

\subsection{White dwarf parameters}

We fitted the upper Balmer line profiles obtained at APO using a grid of model atmospheres in
local thermodynamic equilibrium (LTE).
The model atmospheres include the effect of convection although for this particular model less than $\sim 3$\% of the flux is
carried through convection at $\tau_{\rm R}\approx 1$. More importantly, the models include the
effect of hydrogen line-blanketing that induces cooling of low optical depth layers, and 
stimulates the emergence of a narrow Balmer absorption core clearly seen at H$\alpha$.
Residuals to Balmer line fits are often observed, particularly in line cores, which may stem from non-LTE
effects in hotter atmospheres, or possibly missing opacities.
The new parameters are consistent with our previous analysis and that of \cite{kil2009}.
Table~\ref{tbl-3} summarises the observed properties of the white dwarf in the binary LP~400-22.
We adopted an average of published values of the effective temperature and surface gravity.

\subsection{Binary parameters}

\begin{figure*}
\centering
\includegraphics[width=0.95\columnwidth]{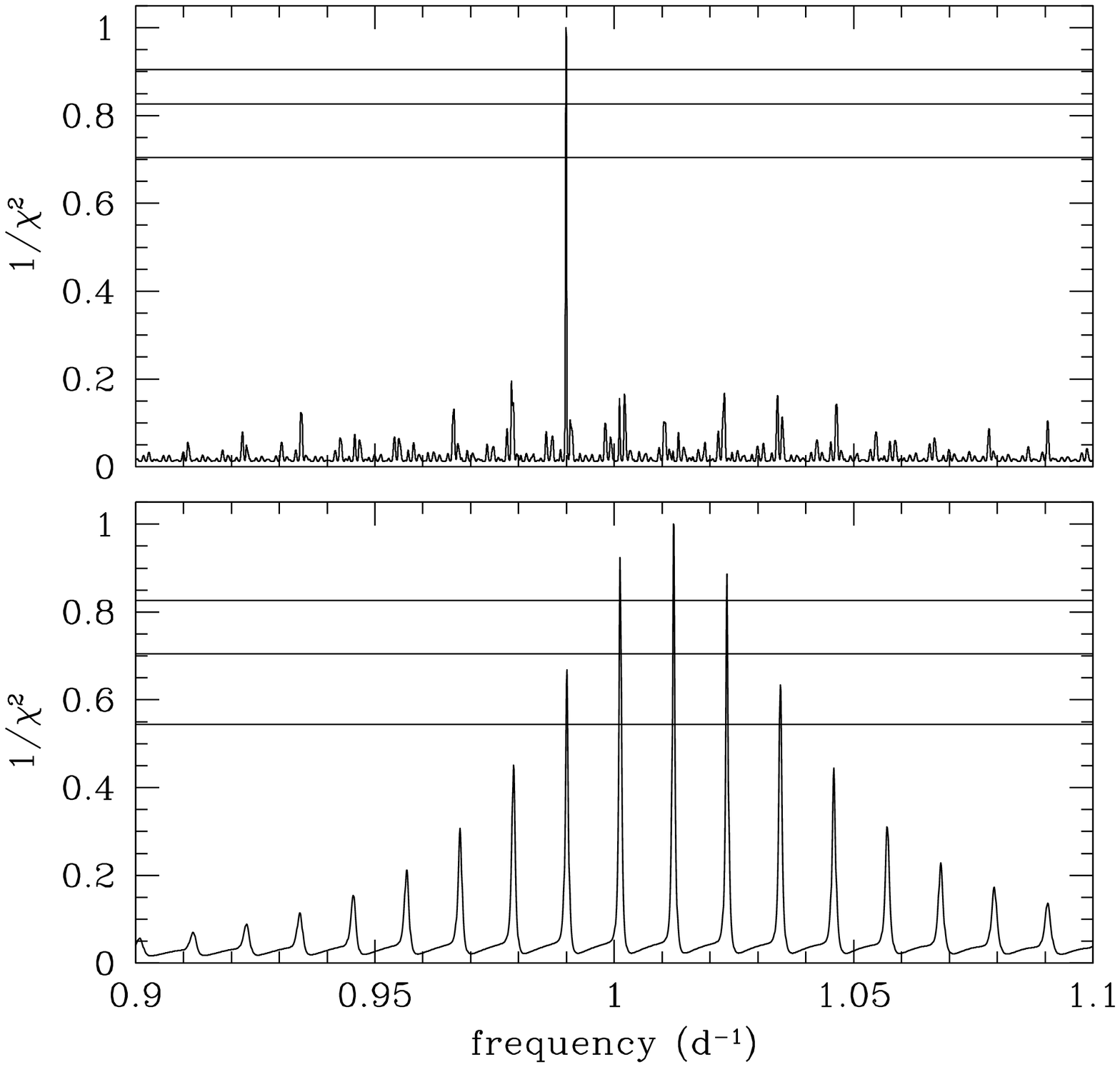}
\includegraphics[width=0.95\columnwidth]{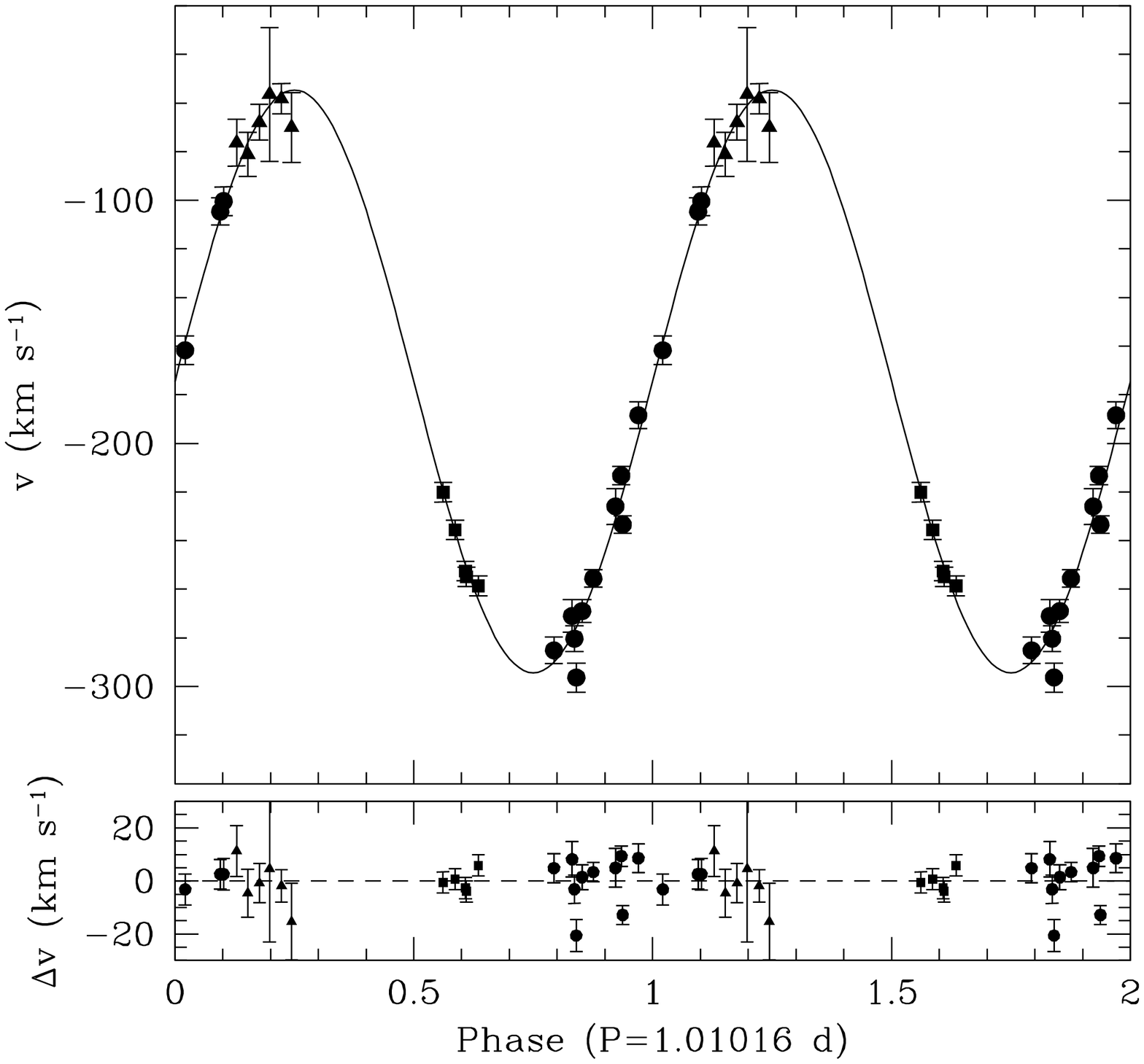}
\caption{({\it Left}) Periodogram of published and new radial velocities of the binary LP~400-22 for ({\it top}) the combined data set (APO, KPNO and \citet{kil2009}), and ({\rm bottom}) for the data set of \citet{kil2009} alone. The analysis shows that the correct period is $P=1.01016$ d. ({\it Right}) Measured velocities and residuals folded on our ephemeris: the triangles are from APO, the squares from KPNO, and the circles from \citet{kil2009}.}
\label{fig2}
\end{figure*}

Figure\ref{fig2} shows the periodograms and best fit radial velocity curve. We fitted the radial velocity measurements, weighted according to Table~\ref{tbl-2}, with
sinusoidal functions while varying the period $P$, the mean and semi-amplitude velocities $\gamma$ and $K$, 
and the initial epoch $T_0$. The corresponding $\chi^2$ values were then normalised to the best-fitting $\chi^2$
and the significance levels  drawn at 66, 90, and 99\%.

We analysed two separate data sets. The first, using the measurements of \citet{kil2009} alone, imply a period of $P=0.9878$ and a minimum $\chi^2=2.5$, but with 4 other periods that cannot be excluded with 99\% confidence. Combining our data with their data, we selected a single period with 99\% confidence and a minimum $\chi^2=2.2$. A solution excluding the less precise data from APO delivers an identical, although less accurate, period
determination.

\begin{table}
\begin{minipage}{\columnwidth}
\caption{Radial velocities}
\label{tbl-2}
\centering
\renewcommand{\footnoterule}{}
\begin{tabular}{llr}
\hline\hline
Observatory & {HJD\footnote{At mid-exposure.} (2450000+)} & $\varv_{\rm helio}$ (km s$^{-1}$) \\
\hline
APO &  4008.64757 &   $-76.3 \pm  9.6$ \\
    &  4008.67122 &   $-81.1 \pm  9.0$ \\
    &  4008.69597 &   $-67.9 \pm  7.4$ \\
    &  4008.71765 &   $-56.4 \pm  27.5$ \\
    &  4008.74278 &   $-58.2 \pm  6.2$ \\
    &  4008.76422 &   $-70.0 \pm  14.4$ \\
KPNO & 4981.93904 &  $-258.6\pm4.0$  \\
     & 4982.92370 &  $-254.9\pm4.0$  \\
     & 4983.88449 &  $-220.1\pm4.0$  \\
     & 4984.92019 &  $-235.6\pm4.0$  \\
     & 4984.94194 &  $-252.5\pm4.0$  \\
\hline
\end{tabular}
\end{minipage}
\end{table}

Based on the joint data set we determined that
\begin{displaymath}
P= 1.01016\pm0.00005\ {\rm d},
\end{displaymath}
\begin{displaymath}
T_0= 2454008.514\pm0.025,   
\end{displaymath}
where $T_0$ is the initial epoch of superior conjunction of the unseen companion ($\Phi=0$).
We determined residuals of 4, 8, and 8 km s$^{-1}$ for the measurements from KPNO, APO, and \citet{kil2009}, respectively.
We subtracted 
the effect of an estimated gravitational redshift of $\gamma_g=\sqrt{GMg}/c \approx 2.5$ km~s$^{-1}$ 
from the apparent systemic velocity $\gamma = -174.5\pm1.3\ {\rm km\ s}^{-1}$  
and determined the systemic velocity:
\begin{displaymath}
\gamma_{\rm sys} = -177.0\pm1.3\ {\rm km\ s}^{-1}.
\end{displaymath}
The projected velocity semi-amplitude of the white dwarf is 
\begin{displaymath}
K = 119.9\pm2.0\ {\rm km\ s}^{-1},
\end{displaymath}
corresponding to a mass function for the unseen companion of
\begin{displaymath}
f= 0.180\pm0.009\ M_\odot.
\end{displaymath}
From the mass function we deduce a minimum mass of $0.41\,M_\odot$ for the companion, and, assuming
a maximum mass of $1.35\,M_\odot$ for a white dwarf companion, the inclination is constrained to
$31 < i < 65^\circ$. A lower inclination is expected for a more massive neutron star companion.
A main-sequence companion of $0.4M_\odot$ with $M_V\approx10$, comparable to LP~400-22, would have 
left the obvious spectroscopic signatures of a red dwarf near H$\alpha$, so that candidacy may be rejected.

Assuming that the white dwarf is tidally locked to its unseen companion, the expected rotational
velocity is $\varv_{\rm rot}\approx 2$ km s$^{-1}$. This effect is not measurable even when using our best data (KPNO) 
at a velocity resolution of 37 km s$^{-1}$.

\subsection{Properties of the white dwarf}

\begin{figure}
\centering
\includegraphics[width=0.95\columnwidth]{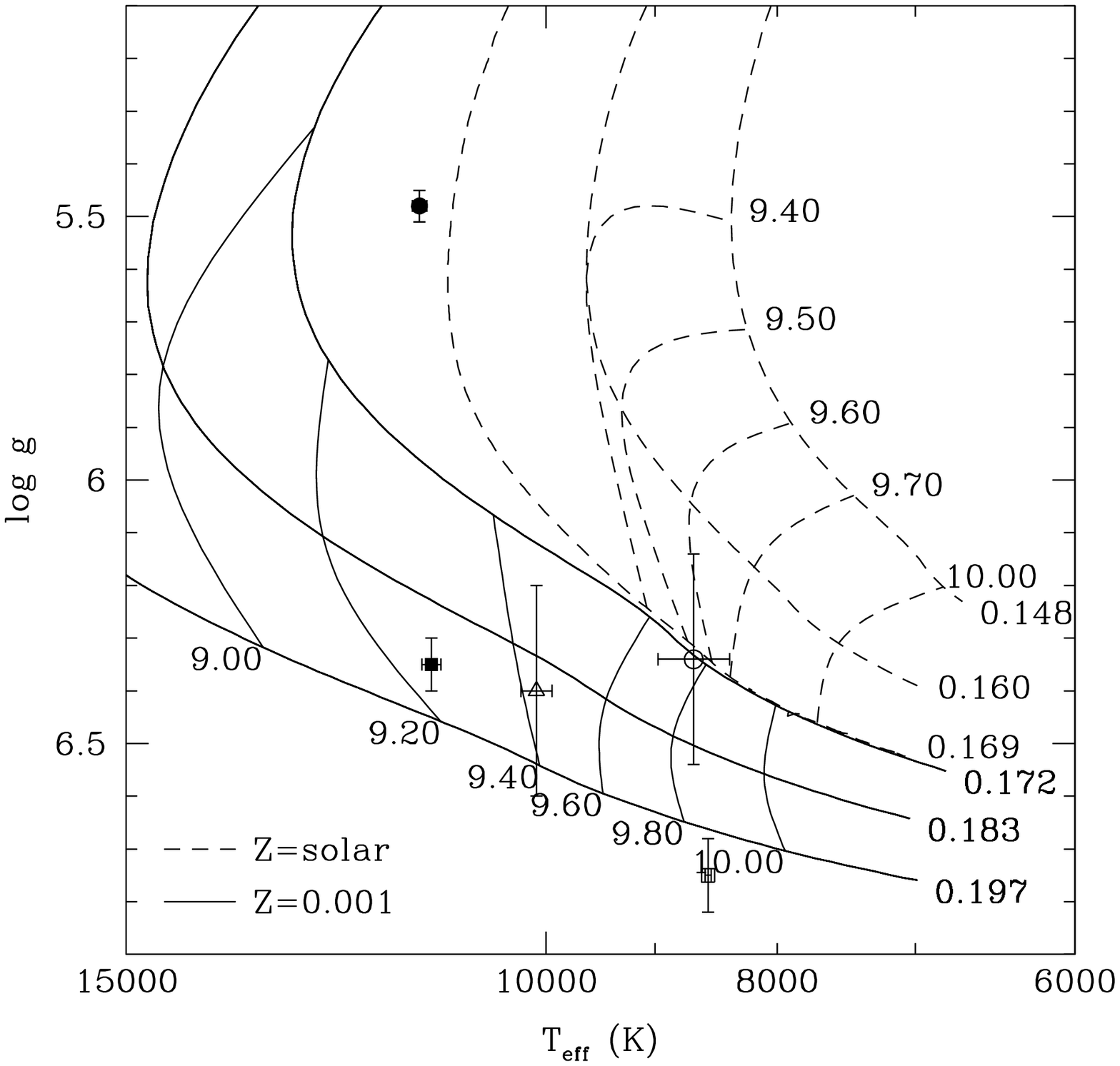} 
\caption{Physical parameters of ELM white dwarfs compared to evolutionary models. The ELM
white dwarfs are LP~400-22 (full square), SDSS~J0917+4638 \citep[full circle,][]{kil2007a,kil2007b},  
and the companions to PSR~J1911-5958A \citep[open triangle][]{bas2006} and PSR~J1012+5307 
\citep[open square or open circle,][respectively]{van1996,cal1998}. The evolutionary models with solar metallicity 
are from \citet{ser2001} and the low-metallicity models ($Z=0.001$) are from \citet{ser2002}.}
\label{fig3}
\end{figure}

Figure~\ref{fig3} compares the observed parameters of LP~400-22 and other ELM white dwarfs with masses below $0.2\,M_\odot$ to the evolutionary models with solar metallicity progenitors \citep{ser2001} and with reduced metallicity \citep{ser2002}.
The isochrones are labelled with $\log{(\rm years)}$ and the curves of constant mass are labelled in units
of $M_\odot$. The effect of an increased metallicity is to reduce the age for a given effective temperature and surface gravity.
The age and mass of LP400-22 derived by \citet{kaw2006} were based on high-metallicity models under 
the assumption of a relatively short total age. If we accept the possibility that the kinematics of
the system is a consequence of old age, then low-metallicity models are appropriate. In this case we re-evaluate the mass at $M=0.19\,M_\odot$ and the age at $t_{\rm cool}=1.8$ Gyr. The corresponding absolute magnitude $M_V=8.98\pm0.12$ mag, from which we infer a distance modulus $m-M=8.24\pm0.14$ and a distance of 444 pc. 
Table~\ref{tbl-3} presents revised properties of the white dwarf. 
Assuming a thick disk or halo origin for the system ($t_{\rm total}\approx8-13$ Gyr) 
implies a pre-RLOF duration, taken as the time for exhaustion of core hydrogen, of 6-11 Gyr.
Based on
evolutionary tracks at $Z=0.001$ of \citet{gir2000}, the mass of the white dwarf progenitor 
is constrained to $0.85\lesssim M\lesssim 1\,M_\odot$. Interestingly,
the case for a low-metallicity progenitor for LP400-22 is supported by the neighbouring case
of PSRJ1911-5958A, which is a probable member of the old cluster NGC~6752 \citep{gra2003} characterised by an age
of 13.8 Gyr and low metallicity $[{\rm Fe/H}]=-1.43$ ($Z=0.001$). With an estimated cooling age
of 2.5 Gyr, a progenitor lifetime of 11.3 Gyr is inferred corresponding to a mass $\sim 0.85\,M_\odot$.

The white dwarfs depicted in Fig.~\ref{fig3} are old, from $\sim$1 to 6 Gyr, and have masses 
narrowly distributed around $0.18\,M_\odot$. Their masses and orbital periods of
14.5 (PSRJ1012+5317), 20.0 (PSRJ1911-5958A), 
and 24.2 hr (LP400-22) provide an anchor for the final mass-period relations for these systems 
\citep{rap1995,nel2004,ben2005}. 
Assuming negligible orbital evolution since the last ROLF episode, the three systems yield
a final mass and period of $(M_f/M_\odot,\log{P_f(d)})=(0.18,-0.1)$. With possibly the lowest mass ($\lesssim0.17\,M_\odot$) of the sample and shortest orbital period (7.6 hr), the white dwarf SDSS~J0917+4638 extends the trend toward shorter
periods.

\section{Discussion}

We show that the binary LP400-22 that comprises an ELM white dwarf
has a period of $1.01016\pm0.00005$ d. Consistent with the kinematics, we adopted low metallicity 
models implying a mass of $0.190\pm0.004$ for the white dwarf and a mass ratio $q=M_{\rm wd}/M_{\rm comp}<0.477$.
The cooling age of the white dwarf is $t_{\rm cool}=1.8\pm0.2$ Gyr. If adopting a total age for the
system of at least 8 Gyr, the progenitor of the white dwarf may have been a late G star with a mass
$\lesssim 1\,M_\odot$.

A possible evolutionary 
scenario initially involved a G+O type binary that subsequently evolved through a 
CE phase which preceded a supernova explosion and that was followed by an episode of 
RLOF from the evolving G star onto the neutron star \citep{wil2002,ben2005}.
In this case, $1.8$ Gyr ago during an RLOF episode, LP400-22 would have appeared as a low-mass
X-ray binary.
Scenarios involving a white dwarf accretor rather than a neutron star
are also possible \citep{nel2001,nel2004}.
Upon certain assumptions concerning the cooling history, \citet{nel2001} successfully produce ELM white dwarfs in close binaries.
An episode of RLOF also implies a past appearance as a cataclysmic variable.

The derived velocity vectors $(U,V,W)$ are not typical of the halo \citep{kaw2006,kil2009}, and it is possible that the Galactic orbit was altered by a supernova
event that added a kick velocity of $\gtrsim 100$ km s$^{-1}$ to the system \citep{hob2005,kie2009}. However, it is unlikely that a supernova event would induce such a large total velocity 
without disrupting the binary in the process.
Therefore, the kinematical properties of this relatively nearby system helps
establish its halo population membership and old age.

\begin{table}
\begin{minipage}{\columnwidth}
\caption{Revised white dwarf parameters}
\label{tbl-3}
\centering
\renewcommand{\footnoterule}{}
\begin{tabular}{llc}
\hline\hline
Parameter  & Measurement & Refs.\footnote{References.--- (1) \citet{kaw2006}; (2) \citet{kil2009}; (3) This work.} \\
\hline
$T_{\rm eff}$  & $11080\pm140$ K &  1 \\ 
               & $11290\pm50$ K &  2 \\
               & $11140\pm90$ K & 3\\ 
$<T_{\rm eff}>$ & $11170\pm90$ K & 3 \\  
$\log{g}$      & $6.32\pm0.08$ (c.g.s.)& 1\\
               & $6.30\pm0.02$ (c.g.s.)& 2\\
	       & $6.42\pm0.06$ (c.g.s.)& 3\\
$<\log{g}>$    & $6.35\pm0.05$ (c.g.s.)& 3\\
$M$\footnote{Mass based on $Z=0.001$ models \citep{ser2002}.} & $0.190\pm0.004\,M_\odot$ & 3 \\
Cooling age    & $1.83\pm0.17$ Gyr & 3 \\
$M_V$          & $8.98\pm0.12$ mag& 3  \\
distance       & $444\pm30$ pc & 3  \\
$(U,V,W)$      & $(-405\pm47,-204\pm13,-30\pm26)$ km s$^{-1}$ & 3   \\
\hline
\end{tabular}
\end{minipage}
\end{table}

\begin{acknowledgements}
S.V. and A.K. are supported by grants IAA300030908 and IAA301630901 from the GA AV, respectively.
Based on observations obtained with the Apache Point Observatory 3.5-meter telescope, which is owned and operated by the Astrophysical Research Consortium.
\end{acknowledgements}

\end{document}